\documentclass[conference]{IEEEtran}
\IEEEoverridecommandlockouts
\usepackage{cite}
\usepackage{amsmath,amssymb,amsfonts}
\usepackage{algorithmic}
\usepackage{graphicx}
\usepackage{textcomp}
\usepackage{xcolor}
\usepackage[T1]{fontenc}
\usepackage[latin9]{inputenc}
\usepackage{geometry}
\usepackage{color}
\usepackage{amsmath}
\usepackage{stmaryrd}
\usepackage{stackrel}
\usepackage{graphicx}
\usepackage{babel}

\def\BibTeX{{\rm B\kern-.05em{\sc i\kern-.025em b}\kern-.08em
   T\kern-.1667em\lower.7ex\hbox{E}\kern-.125emX}}
\begin{document}

\title{A Practical Concatenated Coding Scheme for Noisy Shuffling Channels
with Coset-based Indexing\\
\thanks{This work was funded by the European Union through the ERC Advanced
Grant 101054904: TRANCIDS. Views and opinions expressed are however
those of the authors only and do not necessarily reflect those of
the European Union or the European Research Council Executive Agency.
Neither the European Union nor the granting authority can be held
responsible for them.}}
\author{\IEEEauthorblockN{Javad Haghighat\textsuperscript{1} and Tolga M. Duman\textsuperscript{2}}\IEEEauthorblockA{\textsuperscript{1}\textit{EEE Department, TED University, }Ankara, Turkey, javad.haghighat@tedu.edu.tr \\
\textsuperscript{2}\textit{EEE Department, Bilkent University, }Ankara, Turkey, duman@ee.bilkent.edu.tr}}

\maketitle
\begin{abstract}
Noisy shuffling channels capture the main characteristics of DNA storage
systems where distinct segments of data are received out of order,
after being corrupted by substitution errors. For realistic schemes
with short-length segments, practical indexing and channel coding
strategies are required to restore the order and combat the channel
noise. In this paper, we develop a finite-length concatenated coding
scheme that employs Reed-Solomon (RS) codes as outer codes and polar
codes as inner codes, and utilizes an implicit indexing method based
on cosets of the polar code. We propose a matched decoding method
along with a metric for detecting the index that successfully restores
the order, and correct channel errors at the receiver. Residual errors
that are not corrected by the matched decoder are then corrected by
the outer RS code. We derive analytical approximations for the frame
error rate of the proposed scheme, and also evaluate its performance
through simulations to demonstrate that the proposed implicit indexing
method outperforms explicit indexing.
\end{abstract}

\section{Introduction}

DNA storage systems are receiving significant attention from the research
community, thanks to their longevity and their impressive storage
density \cite{Church2012}-\cite{Heckel2019}. The basic idea in DNA
storage is to employ a synthesizer that takes information bits as
input and maps them to synthetic DNA strands. However, due to technical
limitations in current synthesizing technologies, synthetic strands
are limited to a few hundreds of nucleotides in length. Therefore,
data has to be divided into short segments that are then written on
short strands and are stored in a solution known as the DNA pool. 

The DNA pool has a fundamental disadvantage compared to other storage
environments such as disks and magnetic tapes; that is, DNA pool is
not able to maintain the order of the stored strands (since the strands
are floating in a solution and their physical position cannot be fixed).
Consequently, when the information is being read from the pool, there
is no guarantee that the strands are sequenced (read) in the same
order as they are synthesized (written). In short, the output of the
sequencer is a shuffled version of the strands generated by the synthesizer.
Furthermore, errors are likely to occur during the synthesis, during
the storage, and while sequencing the strands. For this reason, the
end-to-end channel between the original data and the output of the
sequencer may be modeled as a noisy shuffling channel \cite{Shomorony2019}.
The noisy shuffling channel model may be further modified by noting
that the strands are amplified (copied several times) inside the DNA
pool, via the Polymerase Chain Reaction (PCR) process. Hence, the
sequencing process of a randomly selected subset of the stored strands
is more accurately represented by a noisy shuffling-sampling channel
model, in which each strand is sampled a random number of times. These
channel models and their variations are studied in a number of recent
papers including \cite{Shomorony2019}-\cite{Weinberger2022}.

Since the ordering of the segments is not maintained by the noisy
shuffling channel, a mechanism has to be implemented at the transmitter
(i.e., during synthesis) to enable the receiver to restore their order.
The most widely-suggested mechanism is to explicitly assign an index
to each segment, through which the receiver may re-arrange the segments
and restore their order. Although the explicit indexing approach is
proved to be optimal for the asymptotic case \cite{Shomorony2019},
it has several disadvantages in practical finite-length regimes. For
instance, explicit indexing may be prone to errors; i.e., when some
indexes are corrupted by noise, the receiver may mis-detect those
indexes and subsequently may arrange the segments in a (partially)
incorrect order, giving rise to additional errors. This observation
motivates several works including \cite{Lenz2019}-\cite{Weinberger2022}
to focus on error-resilient indexing methods. In \cite{Weinberger2022}
a concatenated coding scheme is proposed for transmission over noisy
shuffling-sampling channels, where the inner code is partitioned into
disjoint sub-codes and each data segment is encoded using a separate
sub-code. It is assumed that a decoder implementation for such an
inner code exists, which provides arbitrarily small decoding error
probabilities. However, only an asymptotic case is considered, where
the number of segments and the segment length grow arbitrarily large,
and no practical implementation is given. 

In this paper, we focus on a case where a sequence of information
bits is sliced into a finite number of short-length segments, and
the segments are transmitted over a noisy shuffling or a noisy shuffling-sampling
channel. We implement a concatenated RS-polar coding scheme, where
each data segment is encoded by a separate coset of a polar code.
No explicit indexing is employed; instead, the decoder decides on
the position of each segment by determining the coset by which that
segment has been encoded. This task is accomplished by aid of a matched
decoding method, where the segment is decoded by all cosets and a
reliability metric is calculated based on which the correct (matched)
coset is detected. Consequently, the matched decoding approach simultaneously
finds the correct position of the segment and corrects the channel
errors on that segment. However, the matched polar decoder is not
able to correct all the channel errors on every segment, hence the
residual errors are then corrected by the outer RS code. 

To identify a proper reliability metric for detecting the matched
coset, we take advantage of the concept of frozen bits in polar codes,
which are set to zero in this paper. The reliabilities of the frozen
bits at the output of a certain decoder may signal whether that decoder
is matched to the input sequence or not. Motivated by this observation,
we introduce a reliability metric in Section \ref{sec:Proposed-Scheme}
and evaluate its performance in Section \ref{sec:Numerical-Results}.
Also, through analytical approximations on the frame error rate (FER)
of the proposed scheme, we demonstrate that the proposed implicit
indexing method achieves lower FERs compared to the explicit indexing
based solutions.

The paper is organized as follows. The system model is given in Section
\ref{sec:System-Model}. In Section \ref{sec:Proposed-Scheme} we
present our proposed concatenated coding scheme. Section \ref{sec:Performance-Analysis}
provides a performance analysis for the proposed scheme through an
approximation derived on the FER. Section \ref{sec:Numerical-Results}
includes numerical results and discussions. Finally, Section \ref{sec:Conclusions}
concludes the paper. 

\section{\label{sec:System-Model}System Model}

\begin{figure}
\centering\includegraphics[scale=0.35]{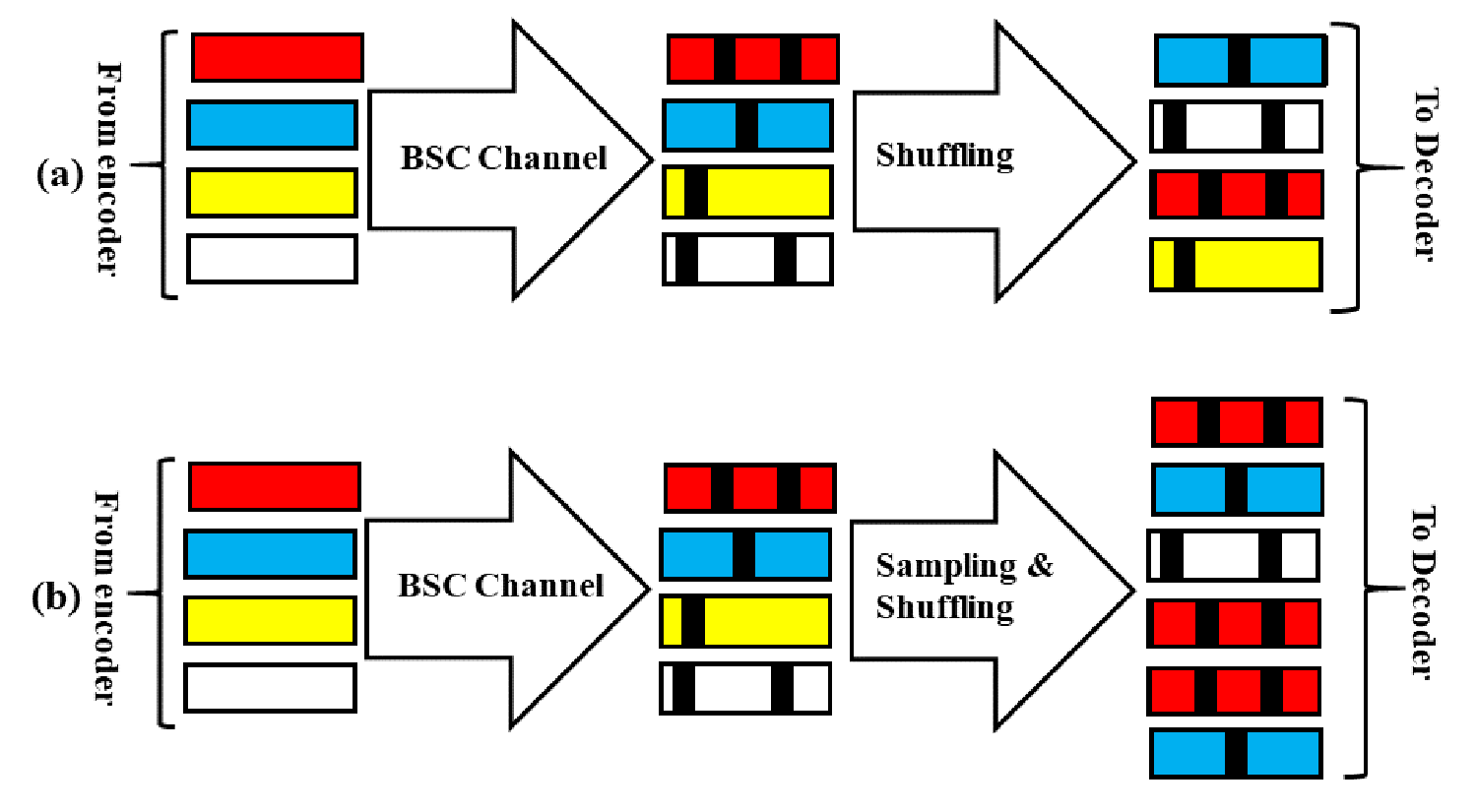}

\vspace{-3mm}

\caption{\label{fig:channel-models}Schematics of (a) noisy shuffling channel
model, and (b) noisy shuffling-sampling channel model. \vspace{-2mm}}

\end{figure}

Schematics of the employed channel models are shown in Fig. \ref{fig:channel-models}.
We assume that $M$ packets with length $L$ (bits) are inputs to
the channel. The input packets are generated by slicing a codeword
of an outer channel code with length $ML$ bits, into $M$ equal-length
segments. The $M$ segments are passed through a noisy channel. Although,
in practice, DNA storage schemes may be affected by several types
of errors, including substitution, deletion and insertion errors,
different works on current DNA storage technologies confirm that substitution
errors are dominant \cite{Goldman2013}-\cite{Heckel2019}. For this
reason and for the sake of simplicity, in this paper, we focus on
noisy channels with substitution errors. Specifically, we consider
a binary symmetric channel (BSC) with crossover probability $\delta$.
For the noisy shuffling channel, the segments at the output of the
BSC are shuffled before being received by the decoder; while for the
noisy shuffling-sampling channel, the segments are sampled $N\geq M$
times with replacement, and the $N$ samples are shuffled before being
received by the decoder (see Fig. \ref{fig:channel-models}). The
ratio $\alpha=\frac{N}{M}$ is called the coverage depth. 

In the specific example depicted in Fig. \ref{fig:channel-models}-b,
$\alpha=1.5$, and the first, the second, the third, and the fourth
packets are sampled $3,2,0$, and $1$ times, respectively. Observe
that in the noisy shuffling-sampling channel model, some segments
may not be sampled, and consequently may not be received by the decoder
at all (e.g., the third segment is not sampled in Fig. \ref{fig:channel-models}-b).
Due to this \textit{missing segments} effect, the expected error rate
experienced over a noisy shuffling-sampling channel is higher compared
to a noisy shuffling channel. However, by increasing the coverage
depth, $\alpha$, the error rate will reduce. 

\section{\label{sec:Proposed-Scheme}Proposed Coding Scheme}

Due to the shuffling process, the order of segments is not preserved
at the output of a noisy shuffling channel. The simplest solution
to this problem is to add indexes to the segments at the encoder.
This approach, which we refer to as \textit{explicit indexing}, enables
the decoder to retrieve the order of the transmitted segments. Although
explicit indexing is a straightforward approach which is widely used
in the literature, it is susceptible to channel noise. Motivated by
this fact and inspired by the work of \cite{Weinberger2022}, in this
section, we propose a concatenated encoding scheme along with an implicit
indexing method that employs different cosets of a polar code to encode
distinct segments; therefore, the position of each segment can be
identified by detecting the coset by which that segment is encoded.
We also propose a matched decoding method that detects the coset corresponding
to a received noisy segment. 

\begin{figure}
\centering\includegraphics[scale=0.3]{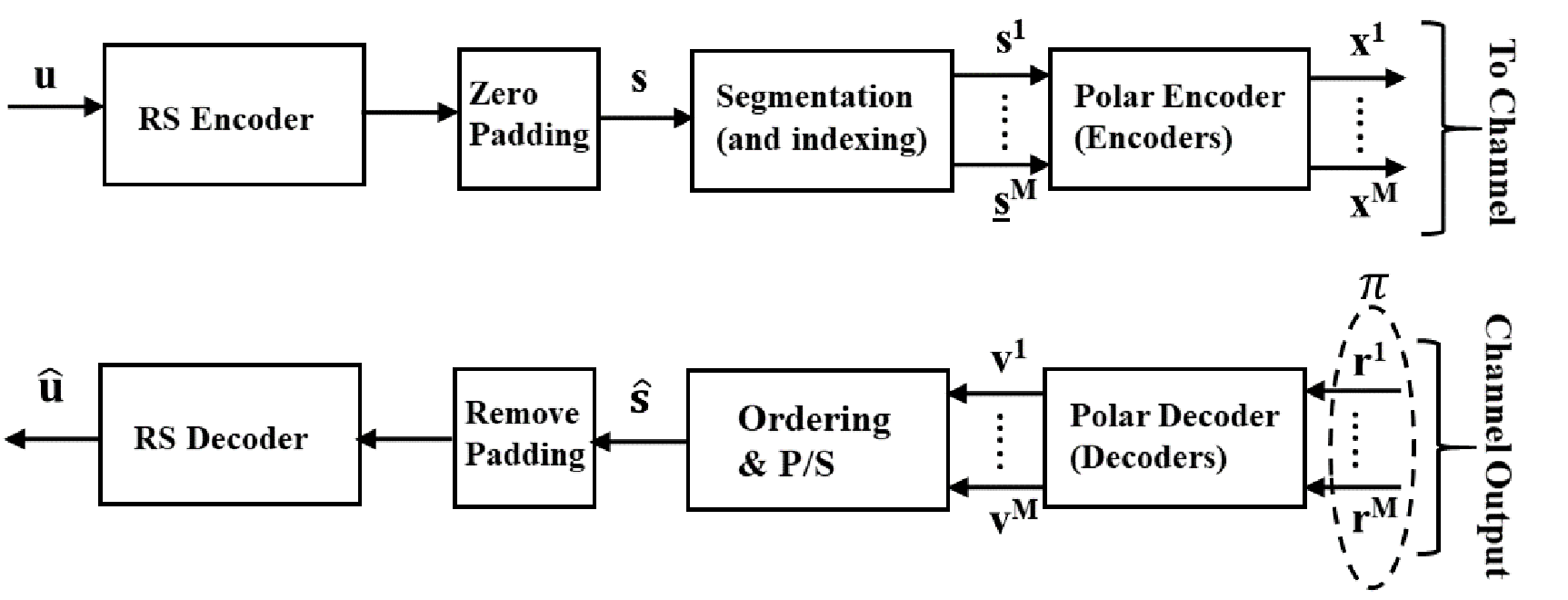}

\caption{\label{fig:Block-diagram-proposed}Block diagram of the proposed encoding
and decoding scheme.}

\vspace{-2mm}
\end{figure}

The block diagram of the proposed scheme is shown in Fig. \ref{fig:Block-diagram-proposed}.
Let $q,k_{o}$ be positive integers. The binary input sequence, $\mathbf{u}$,
consisting of $q\times k_{o}$ bits, is partitioned into $q$-bit
symbols, and then encoded using an $\left(n_{o},k_{o}\right)$ RS
code, where $n_{o}=2^{q}-1$. The RS codeword length is $n_{o}$ symbols,
or equivalently $q\times n_{o}$ bits. This codeword is then zero-padded
by $LM-qn_{o}$ bits to form a binary vector $\mathbf{s}$ with length
$LM$, where $L=\left\lceil \frac{q\times n_{o}}{M}\right\rceil $
and $\left\lceil .\right\rceil $ denotes the ceiling function. The
vector $\mathbf{s}$ is partitioned into $M$ segments of length $L$
bits, denoted by $\mathbf{s}^{1}$ through $\mathbf{s}^{M}$. 

We consider a concatenated coding scheme, where a polar code is applied
as the inner code. We integrate two different indexing implementations
to the polar encoder and the polar decoder blocks of Fig. \ref{fig:Block-diagram-proposed};
namely, explicit indexing, and the proposed coset-based implicit indexing.
In the explicit indexing scheme, an index with length $\left\lceil \log_{2}M\right\rceil $
bits is appended to each segment; hence the length of the indexed
segments is $k_{i}=L+\left\lceil \log_{2}M\right\rceil $ bits. The
indexed segments are encoded by an $\left(n_{i},k_{i}\right)$ polar
code and are transmitted through the channel; i.e., the explicit index
is encoded along with the information bits (the frozen bits are taken
as zero). At the decoder, the received noisy segments are decoded
and the $\left\lceil \log_{2}M\right\rceil $ bits corresponding to
the index are employed to determine the position of each segment in
the entire decoded sequence, $\hat{\mathbf{s}}$. Note that if any
of the $\left\lceil \log_{2}M\right\rceil $ index bits is decoded
with error, the position of the corresponding segment is determined
incorrectly; hence, $\hat{\mathbf{s}}$ is decoded with extra bit
errors introduced by this incorrect ordering process (in addition
to bit errors experienced at the output of a polar decoder). 

In the proposed implicit indexing method, we aim to introduce a scheme
that is more robust to the previously-mentioned index decoding errors.
For this, $M$ cosets of an $\left(n_{i},L\right)$ polar code with
coset leaders $\mathbf{e}^{1}$ through $\mathbf{e}^{M}$ are selected.
For each $m$, $\mathbf{s^{m}}$ is encoded using the $m$-th coset.
This encoding process can be performed by encoding $\mathbf{s^{m}}$
using the polar code, followed by modulo-$2$ addition of $\mathbf{e}^{m}$
to the generated codeword. The symbol $\pi$ in Fig. \ref{fig:Block-diagram-proposed}
denotes a random permutation and aims to emphasize that the noisy
codewords, $\mathbf{r}^{1},\ldots,\mathbf{r}^{M}$, are received out
of order. The decoding process for each received vector, $\mathbf{r}^{m'}$,
$1\leq m'\leq M$ is performed as follows. For all $1\leq m\leq M$,
the vector $\mathbf{r}^{m'}\varoplus\mathbf{e}^{m}$ is fed to a decoder
of the $\left(n_{i},L\right)$ polar code; i.e., $\mathbf{r}^{m'}$
is decoded using the $m$-th coset for all $m$ ($\varoplus$ denotes
bit-wise modulo-$2$ addition). Although the $M$ decoders are identical,
for clarification purposes we denote them by $M$ parallel decoders,
denoted by decoders numbered $1$ through $M.$ After completing the
decoding process, a metric $\xi_{m',m}$ is calculated to quantify
the likelihood of the event that the $m$th decoder is matched to
$\mathbf{r}^{m'}$ (i.e., the event that $\mathbf{r}^{m'}$ is in
fact a noisy copy of a vector which is encoded using the $m$-th coset
of the original code). After completing this process, $\hat{m}=\underset{m}{\mathbf{argmax}}\,\xi_{m',m}$
is determined, and the output of the $\hat{m}$-th decoder is given
as $\mathbf{v}^{m'}$, the decoded version of $\mathbf{r}^{m'}$.
This decoded sequence, $\mathbf{v}^{m'}$, is then written in the
location corresponding to the $\hat{m}$th segment in $\hat{\mathbf{s}}$.
After completing this matched decoding process for all $m'$, the
padded bits are removed from $\hat{\mathbf{s}}$ and the resulting
sequence is decoded by the outer RS decoder to obtain the eventual
decoded bit sequence, $\hat{\mathbf{u}}$. 

In order to find a suitable metric, $\xi_{m',m}$, we propose to exploit
the notion of frozen bits in polar codes. The frozen bits are the
bits that are forced to specific values (e.g., $0$'s in our case)
at the encoder. Each $L$-bit segment, $\mathbf{s}^{m}$ is encoded
by taking an $n_{i}$-bit sequence and filling its $L$ most reliable
positions by the bits of $\mathbf{s}^{m}$. These positions are determined
by a proper channel reliability sequence. The remaining $n_{i}-L$
positions are filled by frozen bits. The resulting $n_{i}$-bit sequence,
denoted by $\tilde{\mathbf{s}}^{m}$, is then transformed into the
$n_{i}$-bit codeword $\mathbf{x}^{m}$ by applying the polar transform.
When a received vector $\mathbf{r}^{m'}$ is decoded using the $m$-th
coset decoder, in addition to producing the hard-decision output,
the decoder is capable of producing a vector $\boldsymbol{\text{\ensuremath{\underbar{\ensuremath{\mathcal{L}}}}}}^{m',m}=\left(\mathcal{L}_{1}^{m',m},\mathcal{L}_{2}^{m',m},\ldots,\mathcal{L}_{n_{i}}^{m',m}\right)$
where $\mathcal{L}_{j}^{m',m}$ denotes the log-likelihood ratio (LLR)
of the $j$th bit of $\tilde{\mathbf{s}}^{m'}$; i.e., $\mathcal{L}_{j}^{m',m}=\log\frac{Pr\left(\tilde{s}_{j}^{m'}=0|m\right)}{Pr\left(\tilde{s}_{j}^{m'}=1|m\right)}$,
where $\tilde{s}_{j}^{m'}$ denotes the $j$th bit of $\tilde{\mathbf{s}}^{m'}$
and the conditioning on ``$m$'' aims to clarify the dependency
of the LLRs on the employed decoder. These LLR values provide a natural
way of measuring the reliability of the decoder output as follows:
Let $\mathcal{F}$ denote the set of indexes of all frozen bits. Define: 

\vspace{-3mm}

\begin{equation}
\xi_{m',m}=\sum_{j\in\mathcal{F}}\mathcal{L}_{j}^{m',m}.\label{eq:metric}
\end{equation}
When the matched decoder is applied, i.e., when $\mathbf{r}^{m'}$
is decoded in the same coset as the one in which it is encoded, $\xi_{m',m}$
is expected to be large. This is due to the fact that frozen bits
are fixed to zero, hence, under matched decoding, their corresponding
LLR values are expected to be high. On the other hand, for mismatched
decoders, there is no such guarantee. Hence, we propose to detect
the index by measuring the metric, $\xi_{m',m}$, for all $m$, and
choosing the value of $m$ that maximizes $\xi_{m',m}$.

\section{\label{sec:Performance-Analysis}Performance Analysis of the Proposed
Scheme}

In this section, we bound the FER of the proposed coset-based scheme
for a noisy shuffling channel under the minimum distance decoding.
Since the RS code has a minimum distance of $n_{o}-k_{o}+1$ symbols,
a minimum distance decoder definitely corrects up to $\frac{n_{o}-k_{o}}{2}$
symbol errors in every RS codeword (selecting $n_{o}-k_{o}$ even).
According to Fig. \ref{fig:Block-diagram-proposed}, if zero-padding
is neglected, the sequence of symbols delivered to the RS decoder
is $\mathbf{\hat{s}}$. Therefore, if a minimum distance decoder is
applied, the probability of observing a frame error is less than or
equal to the probability of observing more than $\frac{n_{o}-k_{o}}{2}$
symbol errors in $\mathbf{\hat{s}}$. Let us consider two events as
follows:

(i) \textit{A mis-detection event (defined as the event where at least
one of the indexes (cosets) is detected incorrectly).} In this case,
at least one segment of $\mathbf{\hat{s}}$ resembles a randomly generated
sequence. This is due to the fact that when the $m$-th decoder is
not matched to $\mathbf{s}^{m}$, the corresponding segment of $\mathbf{\hat{s}}$
either will be filled with another (mismatched) sequence, or is left
without a candidate sequence, in which case without loss of generality
we assume that it is filled with an all-zero sequence. In both cases,
the expected number of symbol errors over the $m$th segment is large
(with a mean value of $\frac{L}{2}$). Therefore, we consider the
worst-case scenario; i.e., when a mis-detection event occurs we assume
that the frame is always decoded erroneously. 

(ii) \textit{The event that all the indexes are detected correctly.}
In such a case, the probability that each bit of $\mathbf{\hat{s}}$
is in error, is equal to the bit error rate (BER) of the polar code.
If the bit error events are assumed to be independent (that is realistic
given an interleaver is employed), since each symbol consists of $q$
consecutive bits, the symbol error rate in such a case can be evaluated
as:

\vspace{-3mm}

\begin{equation}
p_{s}\left(\mathcal{H}\right)=1-\left\{ 1-p_{b}\left(\mathcal{H}\right)\right\} ^{q},\label{eq:PsH}
\end{equation}
where $p_{b}\left(\mathcal{H}\right)$ denotes the BER of the polar
code and $\mathcal{H}$ is a vector specifying the system parameters
(including $n_{i},L,q,M$, $\delta$). 

Let $p_{d}\left(\mathcal{H}\right)$ denote the probability of a mis-detection
event. Then, based on the discussion on cases (i) and (ii), and assuming
the symbol error events are independent, the FER achieved by a minimum
distance decoder can be bounded as:

\vspace{-3mm}

\begin{equation}
\begin{array}{ll}
P_{e}\left(\mathcal{H}\right)\leq & p_{d}\left(\mathcal{H}\right)+\left(1-p_{d}\left(\mathcal{H}\right)\right)\\
 & \times\left(1-\sum_{j=0}^{\frac{n_{o}-k_{o}}{2}}{n_{o} \choose j}p_{s}^{j}\left(\mathcal{H}\right)\left(1-p_{s}\left(\mathcal{H}\right)\right)^{n_{o}-j}\right)
\end{array}\label{eq:Pe_bound}
\end{equation}

To evaluate the right hand side of (\ref{eq:Pe_bound}), one needs
the values (or estimates) of $p_{b}\left(\mathcal{H}\right)$ and
$p_{d}\left(\mathcal{H}\right)$. To estimate $p_{b}\left(\mathcal{H}\right)$,
one may either employ existing bounds on the error probability of
polar codes (e.g., the bounds provided in \cite{Shuval2017}-\cite{Sadra2022}),
or employ Monte-Carlo simulations. 

While the evaluation of $p_{d}\left(\mathcal{H}\right)$ for polar
codes is left for future work, in the following, we derive a bound
on $p_{d}\left(\mathcal{H}\right)$ for a random coding scheme and
a minimum distance decoding approach, explained below, and employ
that bound for approximating the right hand side of (\ref{eq:Pe_bound}). 

Let $\mathcal{C}=\left\{ \mathcal{C}^{1},\ldots,\mathcal{C}^{M}\right\} $
be a set of $M$ random codes with rate $R$. Each code, $\mathcal{C}^{m}$,
contains $2^{nR}$ codewords and each codeword is drawn uniformly
at random (with replacement) from $\mathcal{B}_{n}=\left\{ \mathbf{b}_{1},\ldots,\mathbf{b}_{2^{n}}\right\} $,
the set of all possible realizations of an $n$-bit sequence. For
each codeword $\mathbf{x}\in\mathcal{C}$, let $\mathcal{M}\left(\mathbf{x}\right)$
denote the index of the code to which $\mathbf{x}$ belongs; i.e.,
$\mathcal{M}\left(\mathbf{x}\right)=m$ if and only if $\mathbf{x}\in\mathcal{C}^{m}$.
Also, let

\vspace{-3mm}

\begin{equation}
\hat{\mathbf{x}}=\underset{\mathbf{x}\in\mathcal{C}}{\mathbf{argmin}}\,d_{H}\left(\mathbf{r},\mathbf{x}\right)\label{eq:min_dist_decoder}
\end{equation}
denote the output of a minimum distance decoder, where $d_{H}\left(\mathbf{r},\mathbf{x}\right)$
denotes the Hamming distance between a codeword $\mathbf{x}$ and
the received vector, $\mathbf{r}$. If more than one codeword is at
a minimum distance from $\mathbf{r}$, one of them is selected uniformly
at random and is declared as $\hat{\mathbf{x}}$. 

In order to derive a bound on $p_{d}\left(\mathcal{H}\right)$, we
first derive an upper bound on $Pr\left(\mathcal{M}\left(\mathbf{\hat{x}}\right)\neq\mathcal{M}\left(\mathbf{x}\right)\right)$
for a randomly selected codeword $\mathbf{x}$, and then, we employ
the union bound to derive an upper bound on $p_{d}\left(\mathcal{H}\right)$.
For this purpose, we follow the approach taken in \cite{MacMullan1998}
to find the exact average error probability of a random code ensemble
over a BSC. 

Let us begin by assuming that a fixed codeword $\mathbf{x}_{0}$ is
transmitted over a BSC with channel crossover probability $\delta$,
and the vector $\mathbf{r}=\mathbf{x}_{0}\oplus\text{\ensuremath{\mathbf{z}}}$
is received where $\mathbf{z}$ is the noise vector that is an independent
and identically distributed (i.i.d.) binary sequence with $Pr\left(z_{j}=1\right)=\delta$.
Without loss of generality and for simplicity of notation, let $\mathbf{x}_{0}\in\mathcal{C}^{1}$.
Let $W_{H}\left(\mathbf{z}\right)$ denote the Hamming weight of $\mathbf{z}$.
Since $\mathbf{r}=\mathbf{x}_{0}\oplus\mathbf{z}$, then $d_{H}\left(\mathbf{r},\mathbf{x}_{0}\right)=W_{H}\left(\mathbf{z}\right)=w$.
Note that $\mathbf{x}_{0}\in\mathcal{C}^{1}$; therefore, a sufficient
condition for obtaining $\mathcal{M}\left(\mathbf{\hat{x}}\right)=1$
is that all the codewords in $\cup_{m=2}^{M}\mathcal{C}^{m}$ have
a Hamming distance greater than $w$ from $\mathbf{r}$. Since there
are $2^{nR}\left(M-1\right)$ codewords in $\cup_{m=2}^{M}\mathcal{C}^{m}$
which are realizations of independent and uniformly distributed random
vectors, we have:

\vspace{-4mm}

\begin{equation}
\begin{array}{ll}
Pr\left(\mathcal{M}\left(\mathbf{\hat{x}}\right)=1|W_{H}\left(\mathbf{z}\right)=w\right)\\
\,\,\,\,\,\,\,\,\,\,\,\,\,\,\,\,\,\,\,\,\,\,\,\,\,\,\,\,\,\,\geq\left\{ Pr\left(d_{H}\left(\mathbf{r},\mathbf{x}'\right)>w|W_{H}\left(\mathbf{z}\right)=w\right)\right\} ^{2^{nR}\left(M-1\right)}
\end{array}\label{eq:Pr_Mxhat_cond}
\end{equation}
where $\mathbf{x}'$ is a random vector with a uniform distribution
over $\mathcal{B}_{n}$. 

By applying the law of total probability and by noting that $d_{H}\left(\mathbf{r},\mathbf{x}'\right)=W_{H}\left(\mathbf{z}\oplus\mathbf{x}_{0}\oplus\mathbf{x}'\right)$,
we find: 

\vspace{-3mm}

\begin{equation}
\begin{array}{ll}
Pr\left(\mathcal{M}\left(\mathbf{\hat{x}}\right)=1\right)\geq\sum_{w=0}^{n}Pr\left(W_{H}\left(\mathbf{z}\right)=w\right)\times\\
\,\,\,\,\,\,\,\,\,\,\,\,\,\,\,\,\left\{ Pr\left(W_{H}\left(\mathbf{z}\oplus\mathbf{x}_{0}\oplus\mathbf{x}'\right)>w|W_{H}\left(\mathbf{z}\right)=w\right)\right\} ^{2^{nR}\left(M-1\right)}
\end{array}\label{eq:Pr_Mxhat_bound2}
\end{equation}
Let us define:

\vspace{-3mm}

\begin{equation}
\mathcal{G}\left(\mathbf{z}\right)=\left\{ \mathbf{b}_{j}\in\mathcal{B}_{n}\,s.t.\,W_{H}\left(\mathbf{z}\oplus\mathbf{b}_{j}\right)>W_{H}\left(\mathbf{z}\right)\right\} .\label{eq:gZ_definition}
\end{equation}
Then:

\vspace{-3mm}

\begin{equation}
\begin{array}{r}
Pr\left(W_{H}\left(\mathbf{z}\oplus\mathbf{x}_{0}\oplus\mathbf{x}'\right)>w|W_{H}\left(\mathbf{z}\right)=w\right)\\
=Pr\left(\mathbf{x}_{0}\oplus\mathbf{x}'\in\mathcal{G}\left(\mathbf{z}\right)|W_{H}\left(\mathbf{z}\right)=w\right)
\end{array}\label{eq:Pr_x_xprime_gz}
\end{equation}
If we define $\text{\ensuremath{\mathcal{A}}}_{w}=\left\{ \mathbf{b}_{l}\in\mathcal{B}_{n}\,s.t.\,W_{H}\left(\mathbf{b}_{l}\right)>w\right\} $,
then for every $\mathbf{z}$ with $W_{H}\left(\mathbf{z}\right)=w$,
there exists a one to one mapping between $\mathcal{A}_{w}$ and $\mathcal{G}\left(\mathbf{z}\right)$
as follows. If $\mathbf{b}_{j}\in\mathcal{G}\left(\mathbf{z}\right)$,
then by definition $W_{H}\left(\mathbf{b}_{j}\oplus\mathbf{z}\right)>w$;
i.e., $\mathbf{b}_{j}\oplus\mathbf{z}\in\mathcal{A}_{w}$. Also, if
$\mathbf{b}_{l}\in\mathcal{A}_{w}$, then $W_{H}\left(\mathbf{z}\oplus\left(\mathbf{b}_{l}\oplus\mathbf{\mathbf{z}}\right)\right)=W_{H}\left(\mathbf{b}_{l}\right)>w$;
i.e., $\mathbf{b}_{l}\oplus\mathbf{\mathbf{z}}\in\mathcal{G}\left(\mathbf{z}\right)$.
Therefore, $|\mathcal{G}\left(\mathbf{z}\right)|=|\mathcal{A}_{w}|$,
which gives:

\vspace{-3mm}

\begin{equation}
|\mathcal{G}\left(\mathbf{z}\right)|=2^{n}-\mathcal{N}\left(n,W_{H}\left(\mathbf{z}\right)\right),\label{eq:gZ_cardinality}
\end{equation}
where $\mathcal{N}\left(n,w\right)=\sum_{h=0}^{w}{n \choose h}$.
Notice that all vectors $\mathbf{z}$ with equal Hamming weights have
an identical $|\mathcal{G}\left(\mathbf{z}\right)|$. Also, since
$\mathbf{x}'$ is uniformly distributed over $\mathcal{B}_{n}$, $\mathbf{x}_{0}\oplus\mathbf{x}'$
is uniformly distributed over $\mathcal{B}_{n}$; hence:

\vspace{-4mm}

\begin{equation}
Pr\left(\mathbf{x}_{0}\oplus\mathbf{x}'\in\mathcal{G}\left(\mathbf{z}\right)|W_{H}\left(\mathbf{z}\right)=w\right)=\frac{2^{n}-\mathcal{N}\left(n,w\right)}{2^{n}}.\label{eq:Pr_x0_xprime}
\end{equation}

Using (\ref{eq:Pr_x_xprime_gz}), (\ref{eq:Pr_x0_xprime}), (\ref{eq:Pr_Mxhat_bound2})
and since $Pr\left(W_{H}\left(\mathbf{z}\right)=w\right)={n \choose w}\delta^{w}\left(1-\delta\right)^{n-w}$,
we obtain:

\vspace{-4mm}

\begin{equation}
\begin{array}{l}
Pr\left(\mathcal{M}\left(\mathbf{\hat{x}}\right)=1\right)\geq\\
\,\,\,\,\,\,\,\,\,\,\,\,\,\,\,\sum_{w=0}^{n}{n \choose w}\delta^{w}\left(1-\delta\right)^{n-w}\left(1-2^{-n}\mathcal{N}\left(n,w\right)\right)^{\left(M-1\right)2^{nR}}
\end{array}\label{eq:Pr_Mxhat_bound3}
\end{equation}
For BSCs with crossover probabilities $\delta\ll1$ (e.g., for $\delta<0.05$),
${n \choose w}\delta^{w}\left(1-\delta\right)^{n-w}$ attains its
maximum for small $w$'s (i.e., $w\ll n$). 

To simplify the calculations, we employ the inequality:

\vspace{-4mm}

\begin{equation}
\left(1-a\right)^{n}\geq\left(1-na\right)\times u_{-1}\left(1-na\right),\,\,-1<a<1,\,n\geq1\label{eq:1_minus_an_bound}
\end{equation}
where $u_{-1}\left(.\right)$ is the unit step function. Using (\ref{eq:Pr_Mxhat_bound3}),
(\ref{eq:1_minus_an_bound}), we obtain $Pr\left(\mathcal{M}\left(\mathbf{\hat{x}}\right)\neq1\right)\leq1-f\left(n,R,\delta\right)$
where:

\vspace{-4mm}

\begin{equation}
\begin{array}{l}
f\left(n,R,\delta\right)=\sum_{w=0}^{n}{n \choose w}\delta^{w}\left(1-\delta\right)^{n-w}\times\\
\left(1-2^{-n\left(1-R\right)}\mathcal{N}\left(n,w\right)\right)^{\left(M-1\right)}u_{-1}\left(1-2^{-n\left(1-R\right)}\mathcal{N}\left(n,w\right)\right)
\end{array}\label{eq:f_nRdelta_LB}
\end{equation}
We may take a similar approach to show that $Pr\left(\mathcal{M}\left(\mathbf{\hat{x}}\right)\neq m\right)\leq1-f\left(n,R,\delta\right)$
if a fixed codeword $\mathbf{x}_{0}\in\mathcal{C}^{m}$ is transmitted.
Since in such a case $\mathcal{M}\left(\mathbf{x}_{0}\right)=m$,
we obtain $Pr\left(\mathcal{M}\left(\mathbf{\hat{x}}\right)\neq\mathcal{M}\left(\mathbf{x}_{0}\right)\right)\leq1-f\left(n,R,\delta\right)$
for all $m$; i.e., for any fixed codeword $\mathbf{x}_{0}\in\mathcal{C}$.
Eventually, by observing that $f\left(n,R,\delta\right)$ does not
depend on the choice of $\mathbf{x}_{0}$, we may generalize the upper
bound for a randomly selected codeword $\mathbf{x}\in\mathcal{C}$,
as:

\vspace{-4mm}

\begin{equation}
Pr\left(\mathcal{M}\left(\mathbf{\hat{x}}\right)\neq\mathcal{M}\left(\mathbf{x}\right)\right)\leq1-f\left(n,R,\delta\right).\label{eq:Pr_Mxhat_neq_Mx}
\end{equation}

Now, assume that $M$ randomly selected codewords, $\mathbf{x}^{1}\in\mathcal{C}^{1},\ldots,\mathbf{x}^{M}\in\mathcal{C}^{M}$,
are transmitted over a noisy shuffling channel and a permutation of
their noisy versions, $\mathbf{r}^{1},\ldots,\mathbf{r}^{M}$, is
received at the decoder as shown in Fig. \ref{fig:Block-diagram-proposed}.
By definition, an index detection error occurs if the index of at
least one of these $M$ codewords is detected incorrectly; i.e., 

\vspace{-4mm}

\begin{equation}
\begin{array}{cl}
p_{d}\left(\mathcal{H}\right) & =Pr\left(\stackrel[m=1]{M}{\cup}\left\{ \mathcal{M}\left(\mathbf{\hat{x}}^{m}\right)\neq\mathcal{M}\left(\mathbf{x}^{m}\right)\right\} \right)\\
 & \stackrel{\left(a\right)}{\leq}\sum_{m=1}^{M}Pr\left(\mathcal{M}\left(\mathbf{\hat{x}}^{m}\right)\neq\mathcal{M}\left(\mathbf{x}^{m}\right)\right)\\
 & \stackrel{\left(b\right)}{=}M\times\left(1-f\left(n,R,\delta\right)\right),
\end{array}\label{eq:Pd_H_bound_final}
\end{equation}
where $\mathbf{\hat{x}}^{m}$ is the output of a minimum distance
decoder with input $\mathbf{r}^{m}$; (a) follows from the union bound,
and (b) follows from (\ref{eq:Pr_Mxhat_neq_Mx}). The right hand side
of (\ref{eq:Pd_H_bound_final}) may be replaced in (\ref{eq:Pe_bound})
as an approximation for the FER as follows:

\vspace{-3mm}

\begin{equation}
\begin{array}{l}
P_{e}\left(\mathcal{H}\right)\approx M\left(1-f\left(n,R,\delta\right)\right)+\left(1-M+M\times f\left(n,R,\delta\right)\right)\\
\,\,\,\,\,\,\,\,\,\,\,\,\,\,\,\,\,\,\,\,\,\,\,\,\,\,\times\left(1-\sum_{j=0}^{\frac{n_{o}-k_{o}}{2}}{n_{o} \choose j}p_{s}^{j}\left(\mathcal{H}\right)\left(1-p_{s}\left(\mathcal{H}\right)\right)^{n_{o}-j}\right)
\end{array}\label{eq:FER_approximated}
\end{equation}

Note that we cannot claim that (\ref{eq:FER_approximated}) gives
an upper bound on the FER, since (\ref{eq:Pd_H_bound_final}) is derived
assuming random coding and minimum distance decoding; whereas, in
the proposed scheme, polar codes and their corresponding decoder are
implemented. Nonetheless, implementation of the minimum distance decoder,
which is the optimal decoder for the BSC channel, is prohibitive due
to its exponentially growing complexity with the block length. Also,
in general polar codes have better distance properties compared to
random codes; i.e., it is expected that if a minimum distance decoder
is applied for polar codes, the proposed RS-polar coding scheme would
achieve FER values lower than those suggested by (\ref{eq:Pd_H_bound_final}).
Therefore, (\ref{eq:Pd_H_bound_final}) is useful in the sense that
it provides an insight on achievable FER values of the proposed scheme,
if optimal decoding is applied.

\section{\label{sec:Numerical-Results}Numerical Results}

\begin{figure}
\centering\includegraphics[scale=0.4]{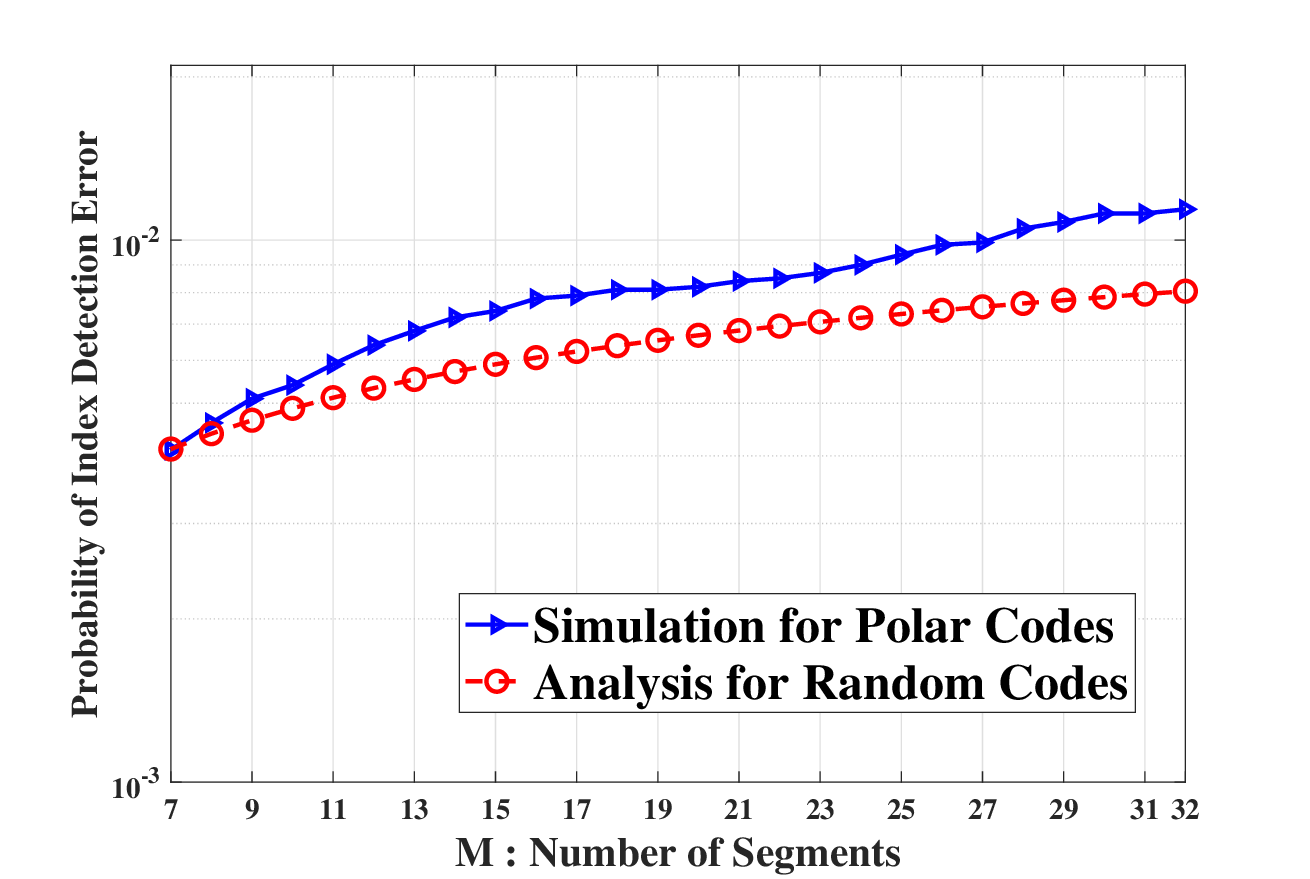}

\caption{\label{fig:Pd-analytic-sim-compare}Comparison between $Pr\left(\mathcal{M}\left(\mathbf{\hat{x}}\right)\protect\neq\mathcal{M}\left(\mathbf{x}\right)\right)$
values found by simulations and by analysis, when $n=n_{i}=128$,
$R=\frac{1}{2}$, $\delta=0.05$.\vspace{-5mm}}
\end{figure}

In this section, we provide numerical examples to quantify the performance
of the proposed scheme, and compare it with that of the explicit indexing
method. Throughout the simulations, we consider a setup with $q=8$,
$n_{o}=255$ symbols, $M=32$, and $n_{i}=128$, unless otherwise
stated. The segment length (without the index bits) is $L=64$ bits.
As explained in Section \ref{sec:Proposed-Scheme}, in the explicit
indexing scheme, each segment is appended by $5$ index bits and the
resulting $69$ bit sequence is encoded by a $\left(128,69\right)$
polar code. In the proposed coset-based scheme, $M=32$ distinct cosets
of a $\left(128,64\right)$ polar code are selected, and the $m$th
segment is encoded by the $m$th coset. We simulate a benchmark scheme,
where for each input bit stream, the $M$ cosets are selected uniformly
at random (but of course, they are assumed to be known at the decoder).
For polar encoding purposes, the most reliable positions are picked
according to the 5G standardization unique channel-reliability sequence
\cite{Bioglio2020}. 

Figure \ref{fig:Pd-analytic-sim-compare} shows the value of $Pr\left(\mathcal{M}\left(\mathbf{\hat{x}}\right)\neq\mathcal{M}\left(\mathbf{x}\right)\right)$
for polar codes found by simulation, compared to the analytical upper
bound derived in (\ref{eq:Pr_Mxhat_neq_Mx}) for random coding and
minimum distance decoding. It is observed that the analysis gives
lower error probabilities; that may be justified by noting that minimum
distance decoding which is the maximum likelihood (ML) decoding over
the BSC is employed for the analysis; i.e., optimal decoding is applied.
Also, there is no proof that the metric introduced in (\ref{eq:metric})
for detecting the matched decoder is optimal (although our empirical
results suggest that it is a good metric). However, we note that unlike
the proposed polar encoding and matched decoding approach, the minimum
distance decoding cannot be implemented in practice, except for very
small block lengths. This is due to the fact that (\ref{eq:min_dist_decoder})
has a computational complexity that grows exponentially with the block
length, $n$ (since the number of codewords is $2^{nR}$). 

\begin{figure}
\centering\includegraphics[scale=0.41]{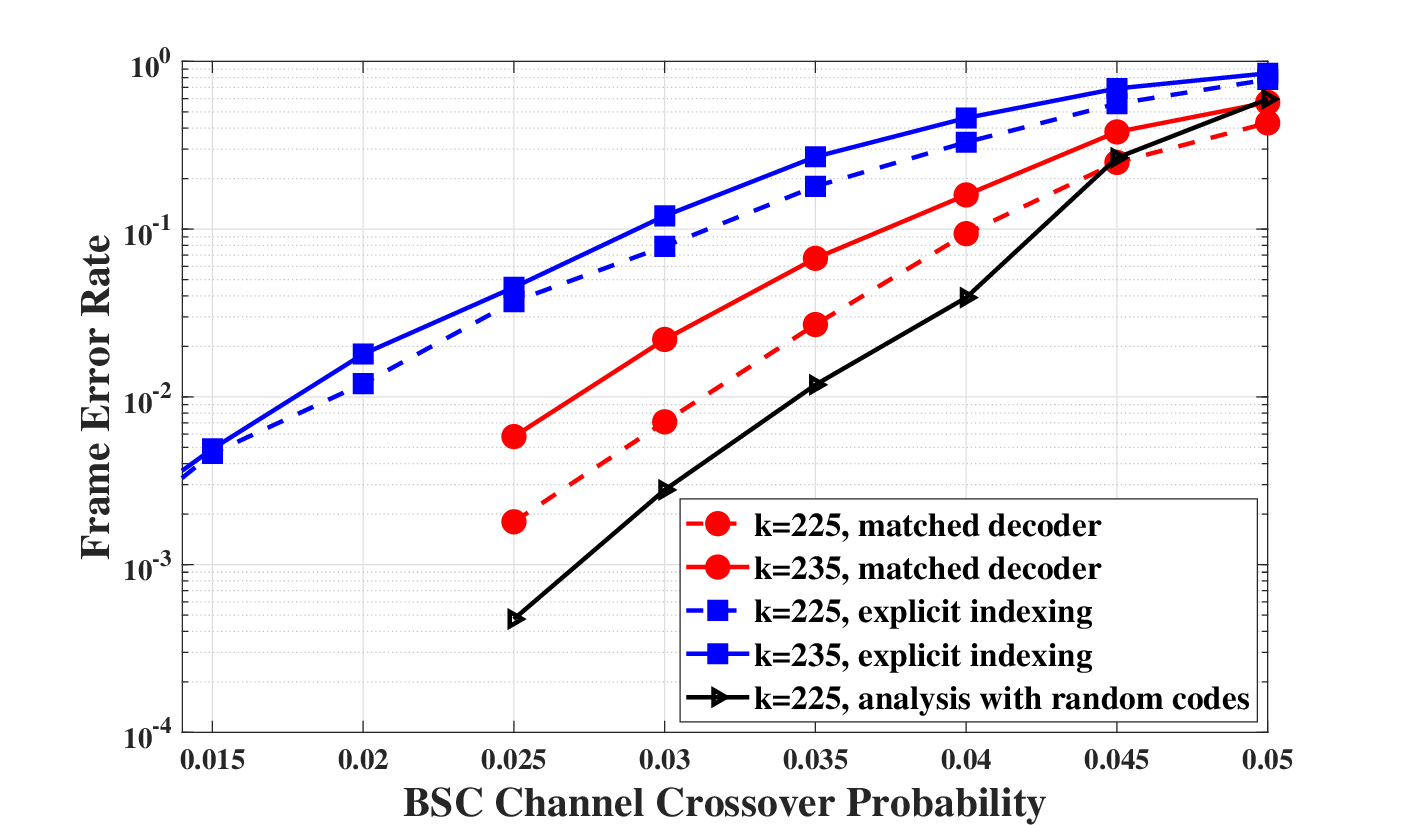}

\caption{\label{fig:FER-perm-no-sample}FERs achieved by explicit indexing
and matched decoding methods for noisy permutation channel, when $n_{0}=255$,
$M=32$, $n_{i}=128$. \vspace{-5mm}}
\end{figure}

Figure \ref{fig:FER-perm-no-sample} shows the FER values found by
simulations for $k_{o}=225$ and $k_{o}=235$. It is observed that
the proposed scheme (labeled as ``matched decoder'') outperforms
the explicit indexing scheme. The gain offered by the proposed scheme
is more significant for case of $k_{o}=225$; also, as expected, the
FER in all cases reduces by decreasing $k_{o}$ at the cost of a reduced
code rate. The black curve shows the analytical result for $k_{o}=225$
(Eq. (\ref{eq:Pd_H_bound_final})). Similar to Fig. \ref{fig:Pd-analytic-sim-compare},
the analysis offers lower FERs compared to the simulation results
based on polar codes (dashed red curve). This observation is justified
by noting that the analytically derived values of $p_{d}\left(\mathcal{H}\right)$
are smaller than the actual values found for polar codes (see Fig.
\ref{fig:Pd-analytic-sim-compare}); hence, using them in (\ref{eq:Pe_bound})
leads to optimistic results. Furthermore, the bound in (\ref{eq:Pe_bound})
is found by applying minimum distance decoding which offers lower
error probabilities compared to practical decoding methods such as
the Berlekamp-Massey algorithm implemented for RS codes in simulations. 

\begin{figure}
\centering\includegraphics[scale=0.41]{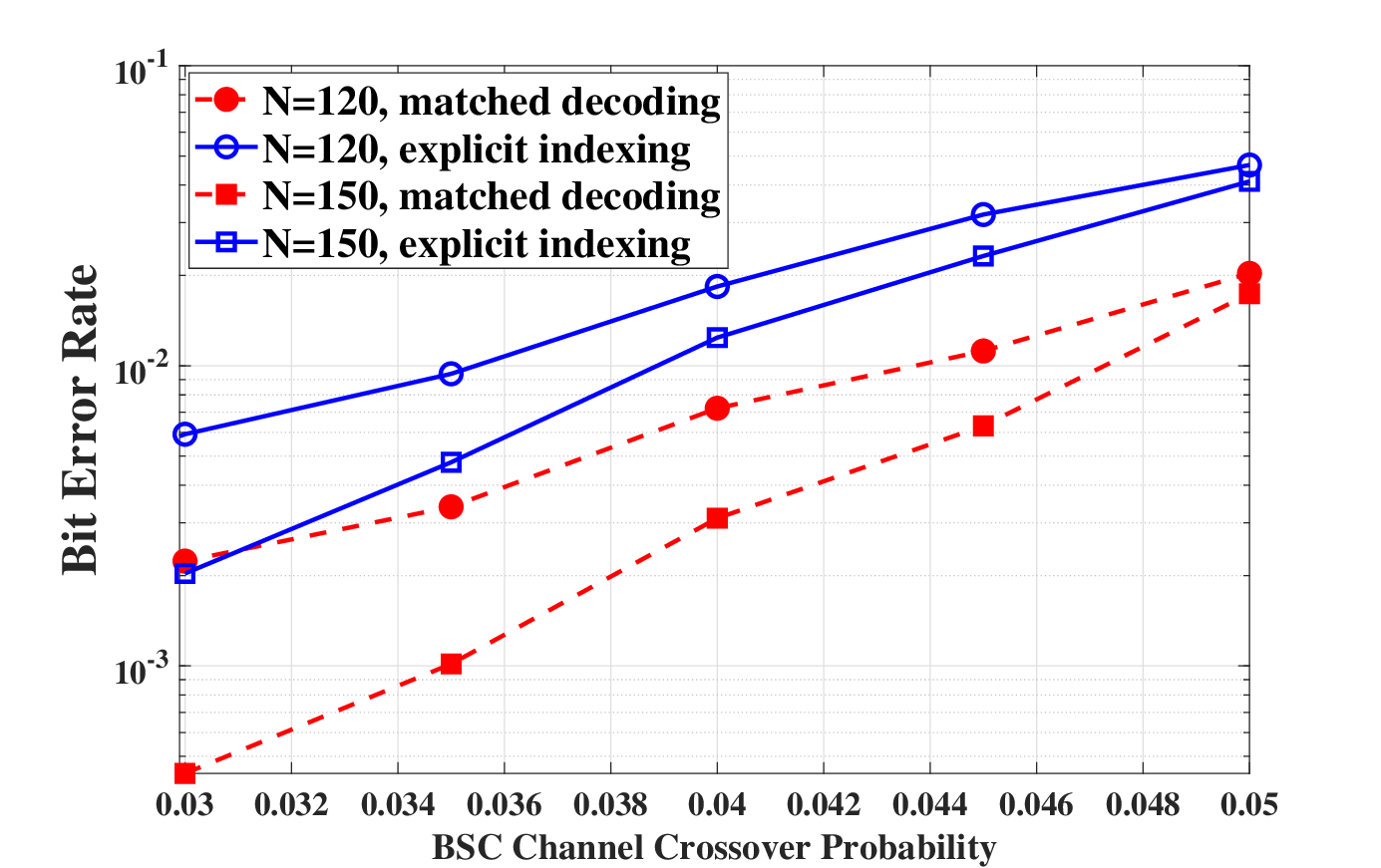}

\caption{\label{fig:BER-shuffling-sampling}BER results for $k_{o}=215$ and
noisy shuffling-sampling channels.\vspace{-5mm}}
\end{figure}

Fig. \ref{fig:BER-shuffling-sampling} shows the bit error rate results
for $k_{o}=215$ and noisy shuffling-sampling channels with $N=120$
and $N=150$ samples. Again, it is observed that the proposed (matched
decoder based) scheme outperforms the explicit indexing scheme. Also,
the BER reduces by increasing the number of samples (i.e., by increasing
the sampling depth $\alpha$). However, this reduction in BER comes
at the cost of larger complexity at the decoder, since more samples
are required to be decoded in order to generate the ordered sequence
$\hat{\mathbf{s}}$. 

\section{\label{sec:Conclusions}Conclusions}

We propose an implicit indexing approach for data transmission over
a noisy shuffling channel, where data is encoded by an outer RS code,
then the RS codeword is sliced into short-length segments, which are
encoded by separate cosets of a polar code. We devise a matched decoding
method that detects the correct coset for each received noisy segment.
We also derive an upper bound for the probability of index detection
error for the case of random codes being employed to encode the $M$
segments. Through this bound, we find an approximation for the FER
of the proposed scheme, which provides insights on the potential performance
of the proposed scheme if optimal decoding is implemented for the
inner code. Performance analysis of the proposed scheme for noisy
shuffling channels with insertion, deletion and substitution errors,
and design of suitable inner codes, are among interesting directions
for future research.

\end{document}